\documentclass[english,reprint]{revtex4-1}
\usepackage[T1]{fontenc}
\usepackage[latin9]{inputenc}
\usepackage{amsmath}
\usepackage{stackrel}
\usepackage{amssymb}
\usepackage{graphicx}
\usepackage{subscript}
\usepackage[unicode=true,pdfusetitle,
 bookmarks=true,bookmarksnumbered=false,bookmarksopen=false,
 breaklinks=false,pdfborder={0 0 1},backref=false,colorlinks=true]
 {hyperref}
\hypersetup{
 linkcolor=blue, citecolor=blue, urlcolor=blue, pdfstartview=XYZ, plainpages=false, pdfpagelabels}
\usepackage{breakurl}

\makeatletter

\AtBeginDocument{\DeclareFontEncoding{T2A}{}{}}

\newcommand{\lyxmathsym}[1]{\ifmmode\begingroup\def\b@ld{bold}
  \text{\ifx\math@version\b@ld\bfseries\fi#1}\endgroup\else#1\fi}

\providecommand{\tabularnewline}{\\}

\@ifundefined{textcolor}{}
{%
 \definecolor{BLACK}{gray}{0}
 \definecolor{WHITE}{gray}{1}
 \definecolor{RED}{rgb}{1,0,0}
 \definecolor{GREEN}{rgb}{0,1,0}
 \definecolor{BLUE}{rgb}{0,0,1}
 \definecolor{CYAN}{cmyk}{1,0,0,0}
 \definecolor{MAGENTA}{cmyk}{0,1,0,0}
 \definecolor{YELLOW}{cmyk}{0,0,1,0}
}

\@ifundefined{date}{}{\date{}}

\providecommand{\tabularnewline}{\\}

\makeatother

\usepackage{babel}
\begin{document}

\title{Stokes vector and its relationship to Discrete Wigner Functions of
multiphoton states}

\author{K.Srinivasan }

\author{G.Raghavan}

\email{gr@igcar.gov.in}

\affiliation{Theoretical studies section, Material Science Group, Indira Gandhi
centre for atomic research, Kalpakkam, Tamilnadu, 603102, India.}

\date{\today}
\begin{abstract}
Stokes vectors and Discrete Wigner functions (DWF) provide two alternate
ways of representing the polarization state of multiphoton systems.
The Stokes vector associated with a n-photon polarization state is
unique, and its Minkowski squared norm provides a direct way of quantifying
entanglement through n-concurrence. However, the quantification of
entanglement from DWF is not straight forward. The DWF associated
with a given quantum state is not unique but depends on the way in
which basis vectors are assigned to various lines in the phase space.
For a Hilbert space of dimension $N$, there exists $N^{N+1}$ such
possible assignments. While a given DWF corresponds to a unique Stokes
vector, the converse is not true. In the present work, we show that,
for each particular assignment called a quantum net, there exist a
unique Hadamard matrix which relates the Stokes vector to the corresponding
DWF. This method provides an elegant and direct method of constructing
the DWFs from the Stokes vector for every possible choice of the quantum
net. Using these results, we derive the relationship between the Stokes
vector of a spin-flipped state and the DWF. Finally, we also present
a method to express the Minkowskian squared norm of the Stokes vector
directly in terms of the DWF. 
\begin{description}
\item [{PACS~numbers}] 03.65.Ta, 03.67.Mn, 42.79.Ta.{\small \par}
\end{description}
\end{abstract}
\maketitle

\section{Introduction}

In quantum optics, the quantum state of multiqubit systems can be
variedly represented through the density matrix, Stokes vector and
Discrete Wigner functions (DWFs). Of these, the density matrix is
by far the most widely used and techniques for entanglement detection
and quantification are defined in terms of this representation \citep{Horodecki19961,PhysRevLett.84.2726,PhysRevLett.80.2245,PhysRevA.61.052306,RevModPhys.81.865,Terhal2000319}.
However, Stokes vectors and the DWFs are equally valid representations
of the state and are both amenable to direct measurements \citep{Berry:77,COLLETT198477,PhysRevA.64.052312,PhysRevA.64.012106}.
Stokes vectors have the advantage that a entanglement measure for
multiqubit state called the generalized concurrence can be defined\citep{PhysRevA.67.032307,PhysRevA.68.022318}.
This measure is basically related to the Minkowski squared norm of
the Stokes vector. Wigner functions are phase space representations
of the state which find extensive applications in quantum optics.
Discrete Wigner functions are real valued and normalized functions
defined over a lattice and find applications in quantum computation,
teleportation and the tomographic reconstruction of qubit systems.
In the present work, we are interested in deriving a direct relationship
between a general $n$-qubit Stokes vector and the DWF, circumventing
the need to compute the density matrix as an intermediate step. In
doing so, we confront the fact that the representation of the state
through the DWF is not unique, but the Stokes vector is. The polarization
state of a photon is represented by the Stokes vector with four parameters,
related to the total intensity and difference in the intensities associated
with measurements using three complementary basis sets. In quantum
mechanical terms, they are related to the expectation values of Pauli
operators with respect to the state $\rho$. DWFs which are discrete
analogues of continuous Wigner functions,\citep{PhysRev.40.749,Hillery1984121}
and have several alternate formulations \citep{Galetti1992513,Cohen1986,ref1,gross1.2393152}.
The present work is based on the construction by Wootters\citep{WOOTTERS19871,Wootters1986,Wootters:2004:PQP:1014615.1014629}
and Gibbons et al\citep{PhysRevA.70.062101}. In Wootters' construction,
the DWF of a $d$-dimensional system is set of real numbers (not necessarily
positive) defined over a $d\times d$ lattice. As will be described
later, the numbers associated with each of the points is obtained
from outcome probabilities of projective measurements using different
basis sets. Since the definition of both Stokes vectors and DWF involve
probabilities of measurement outcomes, the relationship between the
two quantities need careful investigation. Conventionally, the computation
of the Stokes vector of the given DWF involves some arduous calculations.
To appreciate this fact, let us consider the definitions of the DWF
and Stokes vectors in terms of the density matrix : The density matrix
may be defined by $\rho=\sum W_{\alpha}A_{\alpha}$, where $W_{\alpha}$
are the DWF elements and $A_{\alpha}$ are self-adjoint operators
associated with each point of a discrete lattice. For polarization
states of a $n$-photon system, there are $N^{2}$ phase space points,
where $N=2^{n}$, and so the reconstruction of $\rho$ involves the
addition of $N^{2}$ matrices weighted by the DWF element associated
with each point. Once the density matrix is constructed, the corresponding
Stokes vector is calculated using the expression $S_{i_{1}..i_{n}}=\frac{1}{2^{n}}Tr(\rho\sigma_{i_{1}}\otimes\sigma_{i_{2}}\otimes...\otimes\sigma_{i_{n}})$.
Since the density matrix, DWF and the Stokes vector are all related
through linear transformations, such a circuitous procedure can be
avoided if a prescription is provided for computing the Stokes vector
parameters from the DWFs and vice versa. In this article we develop
such a procedure which also addresses the issues related to the choice
of quantum net alluded to earlier. Earlier work of M.Holmes and P.K.Aravind
shows that, for a given DWF there exists a Hadamard matrix which transforms
it to the corresponding Stokes vector, but the form of the Hadamard
transformation for different quantum nets is not provided\citep{Holmes2005}.
To the best of our knowledge, the general prescription for obtaining
this invertible transformation for a $n$-qubit system and its dependence
on the quantum net are absent in the literature. For a discrete phase
space of dimension $N$, there exists $N^{N+1}$ possible quantum
nets. For each quantum net, there is a unique Hadamard matrix that
transforms the DWF to the Stokes vector. Since these Hadamard matrices
are invertible, we show that the DWF can be computed from the Stokes
vector as well. Finally, we discuss some interesting features associated
with the spin flip operation. Bipartite Concurrence and its multipartite
generalizations are important entanglement measures\citep{PhysRevLett.80.2245,PhysRevA.67.032307,PhysRevA.68.022318}.
For $n$-qubit systems, we define a family of Hadamard matrices $S_{H}^{n}$,
such that for each Hadamard matrix $H$, there exists a unique Hadamard
matrix $\tilde{H}$ that takes any given DWF to the Stokes vector
corresponding to the spin flipped state $\tilde{\rho}$. Our article
is arranged as follows: In section II we provide a brief introduction
to the DWF construction of Wootters and Gibbons et al. In section
III-A, for transparency, we illustrate our method for a single qubit
system and generalize the same for a multiqubit system in III-B. In
III-C, we discuss the procedure for obtaining the Stokes vector for
the spin-flipped state. Derivation of Minkowski squared norm in terms
of the DWF is presented in Section IV. We conclude the paper in Section
V with some brief remarks.

\section{Discrete Wigner Function}

There are many constructions available to generalize Wigner functions
to discrete dimensional quantum systems. Here we briefly discuss the
one introduced by Wootters. For quantum systems defined in a Hilbert
space of dimension $N$, the discrete ``phase space'' is a $N\times N$
array of points. Like the continuous case, the horizontal and vertical
axes are associated with two non-commuting observables. For example,
in the discrete array associated with single photon polarization states,
the horizontal and the vertical axes are associated with Pauli's $Z$
and $X$ operators. The points in the phase space are labelled by
the elements of the finite field $\mathcal{F}_{N}$. Since this finite
field exists only for prime or the power of prime dimensions, the
DWF is defined only for such cases. For composite systems, we can
define a basis for the field elements, and express all the elements
of the field as: $q=\underset{i}{\sum}q_{i}e_{i}$, where $q\in\mathcal{F}_{N}$,
$q_{i}\in\mathcal{F}_{r}$ and $e_{i}$ is the element of the basis,
$B=\{e_{1},e_{2},...,e_{n}\}$. Once the basis for the horizontal
axis is fixed, then the basis for the vertical axis can be uniquely
defined. It is easy to see that the discrete phase space has the structure
of an Euclidean space. In this space, we may define a line as a set
of $N$ points subject to the the the equation $aq+bp=c$. Parallel
lines are lines that have the same $a$ and $b$ but different $c$.
Since parallel lines never intersect, they do not have a common point
and non-parallel lines share a single point. There are $N(N+1)$ lines
in discrete phase space which can be grouped into set of $N+1$ parallel
lines. The set of parallel lines are called striations, and each striation
is associated with an observable. The lines in the striations are
associated with the eigenvectors of this observable. We define translational
operators in phase space $\hat{T}_{(\alpha,\beta)}$, whose action
on a line results in translating every point in that line by amount
$(\alpha,\beta)$. In all, there are $N^{2}$ such translational operators
and for a given striation there are $N-1$ translational operators
which leave the lines in the striation invariant. The common eigenvectors
of these $N-1$ translational operators forms a basis which can be
associated with the lines in that striation. In total, $N+1$ orthonormal
basis sets are available which are mutually unbiased. Mutually unbiased
basis sets (MUBs) being defined thus:

Orthonormal basis sets $B_{1}=\left\{ |v_{1}\rangle,|v_{2}\rangle,...,|v_{n}\rangle\right\} $
and $B_{2}=\left\{ |u_{1}\rangle,|u_{2}\rangle,...,|u_{n}\rangle\right\} $
are mutually unbiased if
\[
\left|\langle v_{j}|u_{j'}\rangle\right|^{2}=\frac{1}{N}
\]

Each line is associated with a pure state represented by a rank one
projector $Q(\lambda_{i}^{j})$, where $\lambda_{i}^{j}$ refers to
j$^{th}$ line in the $i$$^{th}$ striation. The sum of DWF elements
along a given line $\lambda_{i}^{j}$ is equal to the expectation
value of the projector $Q(\lambda_{i}^{j})=\left|\lambda_{i}^{j}\right\rangle \left\langle \lambda_{i}^{j}\right|$:
\[
\underset{\alpha\in\lambda}{\sum}W_{\alpha}=Tr(\rho Q(\lambda))
\]
 The discrete Wigner element at the point $\alpha$ is then
\[
W_{\alpha}=\frac{1}{N}\left[\underset{\lambda\ni\alpha}{\sum}Tr(\rho Q(\lambda))-1\right]
\]

This equation can be simplified as 
\begin{equation}
W_{\alpha}=\frac{1}{N}(\rho A_{\alpha})\label{eq:DWF_Def}
\end{equation}

where the $A_{\alpha}$ are the phase space point operators
\begin{equation}
A_{\alpha}=\sum_{\lambda\ni\alpha}Q(\lambda)-I\label{eq:PPoperator}
\end{equation}

These operators $A_{\alpha}$ are self-adjoint and have unit trace
with $Tr(A_{\alpha}A_{\beta})=N\delta_{\alpha\beta}$.

\section{Derivation of results}

\subsection{One qubit system}

Any one qubit system can be represented using the $2\times2$ identity
matrix and the Pauli matrix as the basis
\[
\rho=[s_{0}I+\vec{s}.\vec{\sigma}]
\]
 which may be written as
\begin{equation}
\rho=\stackrel[i=0]{3}{\sum}s_{i}\sigma_{i}\label{eq:DMfPauli}
\end{equation}

Each element of the Stokes vector is given by $s_{i}=\frac{1}{2}Tr(\sigma_{i}\rho)$,
where $\sigma_{i}$'s are the Pauli matrices, $i\in[0,x,y,z]$ The
column vector with coefficients $s_{i}$ as entries is known as the
Stokes vector, i.e., $\vec{S}=(s_{0},s_{1},s_{2},s_{3})^{T}$.

Using the phase space point operators as the basis and DWFs as the
weighting factor one can also express the density matrix as

\begin{equation}
\rho=\underset{\alpha}{\sum}W_{\alpha}A_{\alpha}\label{eq:DMfPP}
\end{equation}

Thus, the Stokes vector and the DWF are characterized by four real
parameters. From Eq (\ref{eq:DMfPauli}) and Eq (\ref{eq:DMfPP}),the
difference between these two representations is the following: Each
element in the Stokes vector is reconstructed by the difference in
the intensities or the probabilities in three mutually unbiased basis
sets. In the case of the DWF, projective measurements yield only the
sum of the DWF elements associated with a line and not the individual
entries. Hence, to obtain the value of a single element, three projective
measurements would be required. 

For a one qubit system, the horizontal and vertical axis are associated
with Pauli's $\sigma_{Z}$ and $\sigma_{X}$ operators respectively.
The finite field elements $\mathcal{F}_{2}=\left\{ 0,1\right\} $
are used to label the points in this discrete $2\times2$ ``phase
space''. Lines in the horizontal axis are associated with the eigenvectors
of the $\sigma_{Z}$ operators denoted by $|H\rangle$ and $|V\rangle$.
Lines in the vertical axis are associated with the eigenvectors of
the $\sigma_{X}$ operators denoted by $|D\rangle$ and $|A\rangle$.
Finally the diagonal lines are associated with the eigenvectors of
the $\sigma_{y}$ operators $|R\rangle$ and $|L\rangle$. The assignment
of these eigenstates to the lines in the phase space is not unique.
Each possible assignment is known as a quantum net. To facilitate
further analysis we now represent the set of Wigner elements $\{W_{00},W_{01},W_{10},W_{11}\}$
by a column vector $W=(W_{00},W_{01},W_{10},W_{11})^{T}$. Denoting
the DWF of the Pauli matrices $\sigma_{0}$, $\sigma_{x}$, $\sigma_{y}$
and $\sigma_{z}$ by $W^{I}$, $W^{X}$, $W^{Y}$ and $W^{Z}$ respectively
the DWFs of the Pauli's operators take the form shown in Table \ref{DWFpauli}.

\begin{table}
\protect\caption{DWFs $W^{I}$, $W^{X}$, $W^{Y}$ and $W^{Z}$ of the $2\times2$
identity matrix and Pauli matrices $\sigma_{X}$, $\sigma_{Y}$, $\sigma_{Z}$
respectively.}

\begin{tabular}{|c|c|c|}
\hline 
$A$ & $\frac{1}{2}$ & $\frac{1}{2}$\tabularnewline
\hline 
$D$ & $\frac{1}{2}$ & $\frac{1}{2}$\tabularnewline
\hline 
$I$ & $H$ & $V$\tabularnewline
\hline 
\end{tabular}$\;\;\;$%
\begin{tabular}{|c|c|c|}
\hline 
$A$ & $-\frac{1}{2}$ & $-\frac{1}{2}$\tabularnewline
\hline 
$D$ & $\frac{1}{2}$ & $\frac{1}{2}$\tabularnewline
\hline 
$X$ & $H$ & $V$\tabularnewline
\hline 
\end{tabular}$\;\;\;$%
\begin{tabular}{|c|c|c|}
\hline 
$A$ & $-\frac{1}{2}$ & $\frac{1}{2}$\tabularnewline
\hline 
$D$ & $\frac{1}{2}$ & $-\frac{1}{2}$\tabularnewline
\hline 
$Y$ & $H$ & $V$\tabularnewline
\hline 
\end{tabular}$\;\;\;$%
\begin{tabular}{|c|c|c|}
\hline 
$A$ & $\frac{1}{2}$ & $-\frac{1}{2}$\tabularnewline
\hline 
$D$ & $\frac{1}{2}$ & $-\frac{1}{2}$\tabularnewline
\hline 
$Z$ & $H$ & $V$\tabularnewline
\hline 
\end{tabular}\label{DWFpauli}
\end{table}

From Table \ref{DWFpauli}, it is clear that each element of the DWF
of the Pauli matrices is $\frac{1}{2}$ multiplied by some phase factor.
If $U$ and $V$ are the DWF of the two operators $\rho_{U}$ and
$\rho_{V}$, then 
\begin{equation}
Tr(\rho_{U}\rho_{V})=2\underset{\alpha}{\sum}U_{\alpha}V_{\alpha}
\end{equation}

Therefore, the Stokes vector $S$ can be represented using this fact
by,

\begin{align}
S_{0}= & \mbox{\ensuremath{\underset{\alpha}{\sum}}}W_{\alpha}^{I}W_{\alpha}\nonumber \\
S_{x}= & \underset{\alpha}{\sum}W_{\alpha}^{X}W_{\alpha}\label{eq:StokeW1}\\
S_{y}= & \mbox{\ensuremath{\underset{\alpha}{\sum}}}W_{\alpha}^{Y}W_{\alpha}\nonumber \\
S_{z}= & \mbox{\ensuremath{\underset{\alpha}{\sum}}}W_{\alpha}^{Z}W_{\alpha}\nonumber 
\end{align}

Using the Eq (\ref{eq:StokeW1})and the DWF of the Pauli operators
from the Table \ref{DWFpauli}, the Stokes vector can be expressed
as

\begin{align}
S_{0}= & \mbox{\ensuremath{\frac{1}{2}}\ensuremath{\underset{\alpha}{\sum}}}W_{\alpha}\nonumber \\
S_{x}= & \mbox{\ensuremath{\frac{1}{2}}\ensuremath{{\displaystyle \sum_{\mbox{ij}}}}\ensuremath{(-1\ensuremath{)^{j}}}}W_{ij}\label{eq:StokesW2}\\
S_{y}= & \mbox{\ensuremath{\frac{1}{2}}\ensuremath{{\displaystyle \sum_{\mbox{ij}}}}\ensuremath{(-1\ensuremath{)^{i}}}}W_{ij}\nonumber \\
S_{z}= & \mbox{\ensuremath{\frac{1}{2}}\ensuremath{{\displaystyle \sum_{\mbox{ij}}}}\ensuremath{(-1\ensuremath{)^{i\oplus j}}}}W_{ij}\nonumber 
\end{align}

where $\oplus$ is addition modulo two. If the DWF is represented
as a column vector, then Eq (\ref{eq:StokesW2}) can be simplified
as
\begin{equation}
\left(\begin{array}{c}
S_{0}\\
S_{x}\\
S_{y}\\
S_{z}
\end{array}\right)=\frac{1}{2}\left(\begin{array}{cccc}
1 & 1 & 1 & 1\\
1 & -1 & 1 & -1\\
1 & -1 & -1 & 1\\
1 & 1 & -1 & -1
\end{array}\right)\left(\begin{array}{c}
W_{00}\\
W_{01}\\
W_{10}\\
W_{11}
\end{array}\right)\label{eq:S_WF}
\end{equation}
\begin{equation}
S=HW\label{eq:S_WF2}
\end{equation}

where $H$ is a constant times a Hadamard matrix. This Hadamard matrix
depends on the choice of the quantum net used to represent the Pauli
operators as given in table-\ref{DWFpauli}. For the one qubit system,
there are $8$ possible quantum nets. So that, for each quantum net,
there is one Hadamard matrix that takes the DWF to the corresponding
Stokes vector. The equation given above can be rewritten as
\begin{equation}
\left(\begin{array}{c}
S_{0}\\
S_{x}\\
S_{y}\\
S_{z}
\end{array}\right)=\frac{1}{2}\left(\begin{array}{c}
W_{00}+W_{01}+W_{10}+W_{11}\\
W_{00}-W_{01}+W_{10}-W_{11}\\
W_{00}-W_{01}-W_{10}+W_{11}\\
W_{00}+W_{01}-W_{10}-W_{11}
\end{array}\right)\label{eq:S_WF3}
\end{equation}

Since the sum of the Wigner elements along a line is associated with
the probabilities Eq (\ref{eq:S_WF3}) can be written as

\begin{equation}
\left(\begin{array}{c}
S_{0}\\
S_{x}\\
S_{y}\\
S_{z}
\end{array}\right)=\frac{1}{2}\left(\begin{array}{c}
1\\
P(+)-P(-)\\
P(R)-P(L)\\
P(H)-P(V)
\end{array}\right)\label{eq:S_WFp}
\end{equation}

This is a well known equation for reconstructing the general polarization
state of the photon using over-complete measurements. The phase factors
in Eq (\ref{eq:StokesW2}) may change for different quantum nets,
however they result in the same probabilities. This transformation
given in Eq (\ref{eq:S_WF}) is invertible and therefore the DWF is
readily constructed from the Stokes vector as 
\begin{equation}
\left(\begin{array}{c}
W_{00}\\
W_{01}\\
W_{10}\\
W_{11}
\end{array}\right)=\frac{1}{2}\left(\begin{array}{cccc}
1 & 1 & 1 & 1\\
1 & -1 & 1 & -1\\
1 & 1 & -1 & -1\\
1 & -1 & -1 & 1
\end{array}\right)\left(\begin{array}{c}
S_{0}\\
S_{x}\\
S_{y}\\
S_{z}
\end{array}\right)\label{eq:W2S1}
\end{equation}
\begin{equation}
W=H^{-1}S\label{eq:W2S2}
\end{equation}

This step is crucial, because when we find the DWF associated with
the given Stokes vector, we should specify what quantum net we are
using for the reconstruction of the DWF. This information about the
quantum net gets embedded in the form of the resulting Hadamard matrix
used for the transformation. In the next section we generalize this
method to $n$-qubit systems. For the $n$ - qubit system there are
$N^{N+1}$ quantum nets with one Hadamard matrix for each quantum
net.

\subsection{N-qubit system}

General $n$-photon polarization states may be described using the
generalized Pauli matrices as basis,
\begin{equation}
\rho=\stackrel[i_{1}...i_{n}=0]{3}{\sum}S_{i_{1}...i_{n}}\sigma_{i_{1}}\otimes\sigma_{i_{2}}\otimes...\otimes\sigma_{i_{n}}
\end{equation}

and the $n$-photon Stokes parameters can be calculated as
\begin{equation}
S_{i_{1}...i_{n}}=\frac{1}{2^{n}}Tr(\rho\sigma_{i_{1}}\otimes\sigma_{i_{2}}\otimes...\otimes\sigma_{i_{n}})
\end{equation}

Let $W$ be the DWF of $\rho$ and $U^{i_{1}...i_{n}}$ be the DWF
of the operator $\sigma_{i_{1}}\otimes\sigma_{i_{2}}\otimes...\otimes\sigma_{i_{n}}$.
Then the Stokes parameters are directly computed from the DWFs of
the $n$-photon polarization state and the generalized Pauli matrices
by
\begin{equation}
S_{i_{1}...i_{n}}=\underset{\alpha}{\sum}W_{\alpha}U_{\alpha}^{i_{1}...i_{n}}\label{eq:Srokes_n}
\end{equation}

The DWF elements of the generalized Pauli matrices are $\pm\frac{1}{2^{n}}$.
Therefore from Eq (\ref{eq:Srokes_n}), the generalized Stokes parameters
can be written using the DWF as
\begin{equation}
S_{i_{1}...i_{n}}=\frac{1}{2^{n}}\underset{\alpha}{\sum}(-1)^{f(\alpha)}W_{\alpha}
\end{equation}

If the elements of the DWF of the $n$-photon polarization state are
arranged as a column vector, then the corresponding Stokes vector
can be calculated using the Hadamard matrices by
\begin{equation}
S=HW\label{eq:StokesM}
\end{equation}

Here, the $N^{2}\times N^{2}$ dimensional Hadamard matrix is weighted
by the factor $\frac{1}{N}$, where $N=2^{n}$. As in the single qubit
case, the inverse of this Hadamard matrix transforms the Stokes parameter
to the corresponding DWF. For the $n$-qubit system, we now define
the set of all Hadamard matrices $S_{H}^{n}$ containing $N^{N+1}$
elements as
\begin{equation}
S_{H}^{n}=\left\{ H(1),\,H(2),\ldots,H(N^{N+1})\right\} 
\end{equation}

where $H(k)$ refers to the Hadamard matrix associated with the $k^{th}$
quantum net.

\subsection{Spin flip operation for $n$-qubit systems}

Spin flip is an important symmetry operation in the fields quantum
information and quantum computation. On a single qubit represented
as a point on the Poincaré sphere, spin flip takes the point to one
anti-podal to it. Since this operation is an involution symmetry operation
involving complex conjugation, it cannot be realized experimentally.
Nevertheless it is an essential tool for entanglement detection and
its quantification. For multiphoton polarization states, the spin
flip operation is defined as $\tilde{\rho}=\sigma_{y}^{\otimes n}\rho^{*}\sigma_{y}^{\otimes n}$,
where the $*$ operation stands for complex conjugation in the computation
basis and $\sigma_{y}$ the Pauli matrix. We may note here that spin
flip is an antiunitary operation\citep{PhysRevLett.83.432}. In a
recent work we proved that the spin flip operation can be performed
on a DWF of the multiqubit systems through a Hadamard transformation
which is independent of the quantum net\citep{sriniGR}. If $W$ and
$\tilde{W}$ are the DWF (arranged as a column vector) of the state
and the spin flipped state respectively, of the $n$-qubit system,
then $\tilde{W}$ can be calculated from $W$ by 
\begin{equation}
\tilde{W}=TW\label{eq:Wtilde-1}
\end{equation}

where $T$ is the Hadamard matrix that is different from one that
used to transform DWF to the Stokes vector. Therefore $T$ does not
belong to the set $S_{H}^{n}$. From Eq (\ref{eq:StokesM}) we can
write the Stokes vector of the spin flipped state as
\begin{equation}
\tilde{S}=H\tilde{W}\label{eq:Stild-1}
\end{equation}

Using Eq (\ref{eq:Wtilde-1}), one can directly calculate $\tilde{S}$
from the given DWF $W$ by
\[
\tilde{S}=HTW
\]

It is important to note here that, $H\in S_{H}^{n}$, but, $T\notin S_{H}^{n}$.
However, the product $HT$ is always the element from the set $S_{H}^{n}$.
We denote this new element by $\tilde{H}=HT$. Interestingly we find,
that for a given $H$(k), $\tilde{H}$(k) is obtained by flipping
each state associated with the quantum net $Q_{i}$. Therefore in
the set $S_{H}$ for every $H$, there exists one unique $\tilde{H}$
which transforms a DWF to its spin flipped Stokes vector.

\section{Minkowsky squared norm of an N-Qubit state in terms of the DWF}

For a $n$ - photon polarization states the multiphoton Stokes parameters
can be defined as follows
\begin{equation}
S_{i_{1}i_{2}...i_{n}}=\frac{1}{2^{n}}Tr(\rho\sigma_{i_{1}}\otimes\sigma_{i_{2}}\otimes...\otimes\sigma_{i_{n}})
\end{equation}

where $\rho$ is the multiqubit density matrix and $\sigma_{i}$'s
are the Pauli matrices. For this Stokes parameter $S_{i_{1}i_{2}...i_{n}}$
we can define a Stokes scalar
\begin{align*}
S_{(n)}^{2}= & \frac{1}{2^{n}}[(S_{0...0})^{2}-\stackrel[k=1]{n}{\sum}\stackrel[i_{k}=1]{3}{\sum}(S_{0...i_{k}...0})^{2}\\
 & -\stackrel[k,l=1]{n}{\sum}\stackrel[i_{k},i_{l}=1]{3}{\sum}(S_{0..i_{k}...i_{l}...0})^{2}\\
 & -...+(-1)^{n}\stackrel[i_{1},i_{2},....,i_{n}=1]{3}{\sum}(S_{i_{1}....,i_{n}})^{2}]
\end{align*}

For a $n$ - photon polarization states this Stokes scalar is invariant
under SLOCC and it is a $O_{0}(1,3)$ group invariant length\citep{PhysRevA.67.032307}.
This Stokes scalar is also called the Minkowskian squared norm. This
Minkowskian squared norm of the Stokes tensor is related to the corresponding
density matrix $\rho$ and its spin flipped density matrix $\tilde{\rho}$
by
\[
S_{(n)}^{2}=Tr(\rho\tilde{\rho})=Tr(R)
\]

where $R=\rho\tilde{\rho}$ is used quantify the entanglement of the
$n$ - qubit systems $S_{(n)}^{2}$. For two qubit systems, concurrence
can be calculated from the eigenvalues of $R$ matrix. Therefore the
quantity $S_{(n)}^{2}=Tr(\rho\tilde{\rho})$ is very useful in calculating
the $n$-concurrence of the multiqubit system. Here, we show that
we can calculate the $n$-concurrence of the system, $C^{2}(|\psi>)=S_{(n)}^{2}$
directly for a given discrete Wigner function. To compute the $n$-concurrence
we use the fact, if $\rho$ and $\sigma$ are two different states
and $W$ and $V$ are the corresponding DWFs, then, $Tr(\rho\sigma)=\underset{\alpha}{N\sum}W_{\alpha}V_{\alpha}$.
So in this case, the pure state concurrence can be written as $C(|\psi>)=\sqrt{Tr(\rho\tilde{\rho})}=\sqrt{N\underset{\alpha}{\sum}W_{\alpha}\tilde{W}_{\alpha}}$.
Using the column vector notation for the DWF, concurrence can be calculated
as
\begin{equation}
C(|\psi>)=\sqrt{NW^{T}\tilde{W}}=\sqrt{NW^{T}HW}\label{eq:concurenceN}
\end{equation}

If $W$ is the DWF of the pure two qubit system, then concurrence
can be calculated directly from the given two qubit DWF by the relation,
$C(|\psi>)=2\sqrt{W^{T}HW}$ .

The relation between the multipartite entanglement measure $S_{(n)}^{2}$,
mixedness of the state $M(\rho)$ and the measure of spin flip symmetry
of the state is given by
\begin{equation}
S_{(n)}^{2}+M(\rho)=I(\rho,\tilde{\rho})\label{eq:sn2}
\end{equation}

where $M(\rho)=1-Tr(\rho^{2})=1-\underset{\alpha}{\sum}W_{\alpha}^{2}$
and $I(\rho,\tilde{\rho})=1-D_{HS}^{2}(\rho-\tilde{\rho})$ can also
be defined as the measure of the indistinguishability of the state
from its spin flipped state, where $D_{HS}^{2}(\rho-\tilde{\rho})=\sqrt{\frac{1}{2}Tr\left[(\rho-\tilde{\rho})^{2}\right]}$
is the Hilbert-Schmidt distance between state and its spin flipped
counterpart\citep{PhysRevA.68.022318}. For a pure multiqubit states,
$M(\rho)=0$, from Eq (\ref{eq:concurenceN}) and Eq (\ref{eq:sn2})
it is clear that 
\begin{equation}
S_{(n)}^{2}=I(\rho,\tilde{\rho})=NW^{T}HW\label{eq:SN2F}
\end{equation}

Therefore for a pure multiqubit systems the entanglement measure $S_{(n)}^{2}$
and $I(\rho,\tilde{\rho})$ are equal and this can be calculated directly
from the multiqubit DWF using Eq (\ref{eq:SN2F}).

\section{Conclusions}

Each of the different representation of the quantum state of multiphoton
systems brings with it certain advantages. Though these representations
are related through linear transformations, the physical insights
and computational advantages provided by one is not readily translated
in terms of the other. For continuous quantum systems, the representation
of the state by Wigner functions provides a clear-cut distinction
between classical and quantum states of light. The Wigner function
for the former are positive but the latter can be negative. Phase-space
representation of the such states provide deep insights into quantum
interference effects. In the case of discrete multiqubit systems,
the relationship between different representations is little explored.
The representation of multiqubit system through density operators
provides, atleast for pure states, the tools to distinguish between
separable and entangled states. Entanglement measures are also defined
in terms of the density operators. In optics, the Stokes vector, both
the classical and quantum versions, provides a direct experimental
means of measurement. The less prevalent DWF too has proved to be
useful in the context of quantum computation, stabilizer codes for
error correction and so forth. Whenever optical qubits are used in
the case of quantum information or quantum computation, it is useful
to understand the relationship between these representations. The
present paper was an attempt at examining such an inter-relationship.
He we have exhibited the existence of a simple relationship between
the DWF and the Stokes vector. The two were shown to be related through
a Hadamard matrix which can be computed for any choice of quantum
net used for the construction of the DWF. Thus, the prescription for
obtaining the complete set $S_{H}$ of Hadamard matrices associated
with every choice of the quantum net was provided.Thus, independent
of the measurement context under which data was obtained, the present
results enable us to easily switch between one representation and
the other. It was further shown that the spin-flipped Stokes vector
can be obtained through a Hadamard matrix which itself member of the
set $S_{H}$. 
\begin{acknowledgments}
One of the authors(K. Srinivasan) acknowledges Indira Gandhi centre
for atomic research, DAE for the award of research fellowship. Useful
suggestions and discussions with Gururaj Kadiri and B.Radhakrishna
is hereby acknowledged by the authors. 
\end{acknowledgments}

\appendix
\bibliographystyle{apsrev4-1}

\begin{thebibliography}{25}%
\makeatletter
\providecommand \@ifxundefined [1]{%
 \@ifx{#1\undefined}
}%
\providecommand \@ifnum [1]{%
 \ifnum #1\expandafter \@firstoftwo
 \else \expandafter \@secondoftwo
 \fi
}%
\providecommand \@ifx [1]{%
 \ifx #1\expandafter \@firstoftwo
 \else \expandafter \@secondoftwo
 \fi
}%
\providecommand \natexlab [1]{#1}%
\providecommand \enquote  [1]{``#1''}%
\providecommand \bibnamefont  [1]{#1}%
\providecommand \bibfnamefont [1]{#1}%
\providecommand \citenamefont [1]{#1}%
\providecommand \href@noop [0]{\@secondoftwo}%
\providecommand \href [0]{\begingroup \@sanitize@url \@href}%
\providecommand \@href[1]{\@@startlink{#1}\@@href}%
\providecommand \@@href[1]{\endgroup#1\@@endlink}%
\providecommand \@sanitize@url [0]{\catcode `\\12\catcode `\$12\catcode
  `\&12\catcode `\#12\catcode `\^12\catcode `\_12\catcode `\%12\relax}%
\providecommand \@@startlink[1]{}%
\providecommand \@@endlink[0]{}%
\providecommand \url  [0]{\begingroup\@sanitize@url \@url }%
\providecommand \@url [1]{\endgroup\@href {#1}{\urlprefix }}%
\providecommand \urlprefix  [0]{URL }%
\providecommand \Eprint [0]{\href }%
\providecommand \doibase [0]{http://dx.doi.org/}%
\providecommand \selectlanguage [0]{\@gobble}%
\providecommand \bibinfo  [0]{\@secondoftwo}%
\providecommand \bibfield  [0]{\@secondoftwo}%
\providecommand \translation [1]{[#1]}%
\providecommand \BibitemOpen [0]{}%
\providecommand \bibitemStop [0]{}%
\providecommand \bibitemNoStop [0]{.\EOS\space}%
\providecommand \EOS [0]{\spacefactor3000\relax}%
\providecommand \BibitemShut  [1]{\csname bibitem#1\endcsname}%
\let\auto@bib@innerbib\@empty
\bibitem [{\citenamefont {Horodecki}\ \emph {et~al.}(1996)\citenamefont
  {Horodecki}, \citenamefont {Horodecki},\ and\ \citenamefont
  {Horodecki}}]{Horodecki19961}%
  \BibitemOpen
  \bibfield  {author} {\bibinfo {author} {\bibfnamefont {M.}~\bibnamefont
  {Horodecki}}, \bibinfo {author} {\bibfnamefont {P.}~\bibnamefont
  {Horodecki}}, \ and\ \bibinfo {author} {\bibfnamefont {R.}~\bibnamefont
  {Horodecki}},\ }\href {\doibase
  http://dx.doi.org/10.1016/S0375-9601(96)00706-2} {\bibfield  {journal}
  {\bibinfo  {journal} {Physics Letters A}\ }\textbf {\bibinfo {volume}
  {223}},\ \bibinfo {pages} {1 } (\bibinfo {year} {1996})}\BibitemShut
  {NoStop}%
\bibitem [{\citenamefont {Simon}(2000)}]{PhysRevLett.84.2726}%
  \BibitemOpen
  \bibfield  {author} {\bibinfo {author} {\bibfnamefont {R.}~\bibnamefont
  {Simon}},\ }\href {\doibase 10.1103/PhysRevLett.84.2726} {\bibfield
  {journal} {\bibinfo  {journal} {Phys. Rev. Lett.}\ }\textbf {\bibinfo
  {volume} {84}},\ \bibinfo {pages} {2726} (\bibinfo {year}
  {2000})}\BibitemShut {NoStop}%
\bibitem [{\citenamefont {Wootters}(1998)}]{PhysRevLett.80.2245}%
  \BibitemOpen
  \bibfield  {author} {\bibinfo {author} {\bibfnamefont {W.~K.}\ \bibnamefont
  {Wootters}},\ }\href {\doibase 10.1103/PhysRevLett.80.2245} {\bibfield
  {journal} {\bibinfo  {journal} {Phys. Rev. Lett.}\ }\textbf {\bibinfo
  {volume} {80}},\ \bibinfo {pages} {2245} (\bibinfo {year}
  {1998})}\BibitemShut {NoStop}%
\bibitem [{\citenamefont {Coffman}\ \emph {et~al.}(2000)\citenamefont
  {Coffman}, \citenamefont {Kundu},\ and\ \citenamefont
  {Wootters}}]{PhysRevA.61.052306}%
  \BibitemOpen
  \bibfield  {author} {\bibinfo {author} {\bibfnamefont {V.}~\bibnamefont
  {Coffman}}, \bibinfo {author} {\bibfnamefont {J.}~\bibnamefont {Kundu}}, \
  and\ \bibinfo {author} {\bibfnamefont {W.~K.}\ \bibnamefont {Wootters}},\
  }\href {\doibase 10.1103/PhysRevA.61.052306} {\bibfield  {journal} {\bibinfo
  {journal} {Phys. Rev. A}\ }\textbf {\bibinfo {volume} {61}},\ \bibinfo
  {pages} {052306} (\bibinfo {year} {2000})}\BibitemShut {NoStop}%
\bibitem [{\citenamefont {Horodecki}\ \emph {et~al.}(2009)\citenamefont
  {Horodecki}, \citenamefont {Horodecki}, \citenamefont {Horodecki},\ and\
  \citenamefont {Horodecki}}]{RevModPhys.81.865}%
  \BibitemOpen
  \bibfield  {author} {\bibinfo {author} {\bibfnamefont {R.}~\bibnamefont
  {Horodecki}}, \bibinfo {author} {\bibfnamefont {P.}~\bibnamefont
  {Horodecki}}, \bibinfo {author} {\bibfnamefont {M.}~\bibnamefont
  {Horodecki}}, \ and\ \bibinfo {author} {\bibfnamefont {K.}~\bibnamefont
  {Horodecki}},\ }\href {\doibase 10.1103/RevModPhys.81.865} {\bibfield
  {journal} {\bibinfo  {journal} {Rev. Mod. Phys.}\ }\textbf {\bibinfo {volume}
  {81}},\ \bibinfo {pages} {865} (\bibinfo {year} {2009})}\BibitemShut
  {NoStop}%
\bibitem [{\citenamefont {Terhal}(2000)}]{Terhal2000319}%
  \BibitemOpen
  \bibfield  {author} {\bibinfo {author} {\bibfnamefont {B.~M.}\ \bibnamefont
  {Terhal}},\ }\href {\doibase http://dx.doi.org/10.1016/S0375-9601(00)00401-1}
  {\bibfield  {journal} {\bibinfo  {journal} {Physics Letters A}\ }\textbf
  {\bibinfo {volume} {271}},\ \bibinfo {pages} {319 } (\bibinfo {year}
  {2000})}\BibitemShut {NoStop}%
\bibitem [{\citenamefont {Berry}\ \emph {et~al.}(1977)\citenamefont {Berry},
  \citenamefont {Gabrielse},\ and\ \citenamefont {Livingston}}]{Berry:77}%
  \BibitemOpen
  \bibfield  {author} {\bibinfo {author} {\bibfnamefont {H.~G.}\ \bibnamefont
  {Berry}}, \bibinfo {author} {\bibfnamefont {G.}~\bibnamefont {Gabrielse}}, \
  and\ \bibinfo {author} {\bibfnamefont {A.~E.}\ \bibnamefont {Livingston}},\
  }\href {\doibase 10.1364/AO.16.003200} {\bibfield  {journal} {\bibinfo
  {journal} {Appl. Opt.}\ }\textbf {\bibinfo {volume} {16}},\ \bibinfo {pages}
  {3200} (\bibinfo {year} {1977})}\BibitemShut {NoStop}%
\bibitem [{\citenamefont {Collett}(1984)}]{COLLETT198477}%
  \BibitemOpen
  \bibfield  {author} {\bibinfo {author} {\bibfnamefont {E.}~\bibnamefont
  {Collett}},\ }\href {\doibase http://dx.doi.org/10.1016/0030-4018(84)90286-4}
  {\bibfield  {journal} {\bibinfo  {journal} {Optics Communications}\ }\textbf
  {\bibinfo {volume} {52}},\ \bibinfo {pages} {77 } (\bibinfo {year}
  {1984})}\BibitemShut {NoStop}%
\bibitem [{\citenamefont {James}\ \emph {et~al.}(2001)\citenamefont {James},
  \citenamefont {Kwiat}, \citenamefont {Munro},\ and\ \citenamefont
  {White}}]{PhysRevA.64.052312}%
  \BibitemOpen
  \bibfield  {author} {\bibinfo {author} {\bibfnamefont {D.~F.~V.}\
  \bibnamefont {James}}, \bibinfo {author} {\bibfnamefont {P.~G.}\ \bibnamefont
  {Kwiat}}, \bibinfo {author} {\bibfnamefont {W.~J.}\ \bibnamefont {Munro}}, \
  and\ \bibinfo {author} {\bibfnamefont {A.~G.}\ \bibnamefont {White}},\ }\href
  {\doibase 10.1103/PhysRevA.64.052312} {\bibfield  {journal} {\bibinfo
  {journal} {Phys. Rev. A}\ }\textbf {\bibinfo {volume} {64}},\ \bibinfo
  {pages} {052312} (\bibinfo {year} {2001})}\BibitemShut {NoStop}%
\bibitem [{\citenamefont {Asplund}\ and\ \citenamefont
  {Bj\"ork}(2001)}]{PhysRevA.64.012106}%
  \BibitemOpen
  \bibfield  {author} {\bibinfo {author} {\bibfnamefont {R.}~\bibnamefont
  {Asplund}}\ and\ \bibinfo {author} {\bibfnamefont {G.}~\bibnamefont
  {Bj\"ork}},\ }\href {\doibase 10.1103/PhysRevA.64.012106} {\bibfield
  {journal} {\bibinfo  {journal} {Phys. Rev. A}\ }\textbf {\bibinfo {volume}
  {64}},\ \bibinfo {pages} {012106} (\bibinfo {year} {2001})}\BibitemShut
  {NoStop}%
\bibitem [{\citenamefont {Jaeger}\ \emph
  {et~al.}(2003{\natexlab{a}})\citenamefont {Jaeger}, \citenamefont
  {Teodorescu-Frumosu}, \citenamefont {Sergienko}, \citenamefont {Saleh},\ and\
  \citenamefont {Teich}}]{PhysRevA.67.032307}%
  \BibitemOpen
  \bibfield  {author} {\bibinfo {author} {\bibfnamefont {G.}~\bibnamefont
  {Jaeger}}, \bibinfo {author} {\bibfnamefont {M.}~\bibnamefont
  {Teodorescu-Frumosu}}, \bibinfo {author} {\bibfnamefont {A.}~\bibnamefont
  {Sergienko}}, \bibinfo {author} {\bibfnamefont {B.~E.~A.}\ \bibnamefont
  {Saleh}}, \ and\ \bibinfo {author} {\bibfnamefont {M.~C.}\ \bibnamefont
  {Teich}},\ }\href {\doibase 10.1103/PhysRevA.67.032307} {\bibfield  {journal}
  {\bibinfo  {journal} {Phys. Rev. A}\ }\textbf {\bibinfo {volume} {67}},\
  \bibinfo {pages} {032307} (\bibinfo {year} {2003}{\natexlab{a}})}\BibitemShut
  {NoStop}%
\bibitem [{\citenamefont {Jaeger}\ \emph
  {et~al.}(2003{\natexlab{b}})\citenamefont {Jaeger}, \citenamefont
  {Sergienko}, \citenamefont {Saleh},\ and\ \citenamefont
  {Teich}}]{PhysRevA.68.022318}%
  \BibitemOpen
  \bibfield  {author} {\bibinfo {author} {\bibfnamefont {G.}~\bibnamefont
  {Jaeger}}, \bibinfo {author} {\bibfnamefont {A.~V.}\ \bibnamefont
  {Sergienko}}, \bibinfo {author} {\bibfnamefont {B.~E.~A.}\ \bibnamefont
  {Saleh}}, \ and\ \bibinfo {author} {\bibfnamefont {M.~C.}\ \bibnamefont
  {Teich}},\ }\href {\doibase 10.1103/PhysRevA.68.022318} {\bibfield  {journal}
  {\bibinfo  {journal} {Phys. Rev. A}\ }\textbf {\bibinfo {volume} {68}},\
  \bibinfo {pages} {022318} (\bibinfo {year} {2003}{\natexlab{b}})}\BibitemShut
  {NoStop}%
\bibitem [{\citenamefont {Wigner}(1932)}]{PhysRev.40.749}%
  \BibitemOpen
  \bibfield  {author} {\bibinfo {author} {\bibfnamefont {E.}~\bibnamefont
  {Wigner}},\ }\href {\doibase 10.1103/PhysRev.40.749} {\bibfield  {journal}
  {\bibinfo  {journal} {Phys. Rev.}\ }\textbf {\bibinfo {volume} {40}},\
  \bibinfo {pages} {749} (\bibinfo {year} {1932})}\BibitemShut {NoStop}%
\bibitem [{\citenamefont {Hillery}\ \emph {et~al.}(1984)\citenamefont
  {Hillery}, \citenamefont {O'Connell}, \citenamefont {Scully},\ and\
  \citenamefont {Wigner}}]{Hillery1984121}%
  \BibitemOpen
  \bibfield  {author} {\bibinfo {author} {\bibfnamefont {M.}~\bibnamefont
  {Hillery}}, \bibinfo {author} {\bibfnamefont {R.}~\bibnamefont {O'Connell}},
  \bibinfo {author} {\bibfnamefont {M.}~\bibnamefont {Scully}}, \ and\ \bibinfo
  {author} {\bibfnamefont {E.}~\bibnamefont {Wigner}},\ }\href {\doibase
  http://dx.doi.org/10.1016/0370-1573(84)90160-1} {\bibfield  {journal}
  {\bibinfo  {journal} {Physics Reports}\ }\textbf {\bibinfo {volume} {106}},\
  \bibinfo {pages} {121 } (\bibinfo {year} {1984})}\BibitemShut {NoStop}%
\bibitem [{\citenamefont {Galetti}\ and\ \citenamefont
  {Piza}(1992)}]{Galetti1992513}%
  \BibitemOpen
  \bibfield  {author} {\bibinfo {author} {\bibfnamefont {D.}~\bibnamefont
  {Galetti}}\ and\ \bibinfo {author} {\bibfnamefont {A.~T.}\ \bibnamefont
  {Piza}},\ }\href {\doibase http://dx.doi.org/10.1016/0378-4371(92)90213-A}
  {\bibfield  {journal} {\bibinfo  {journal} {Physica A: Statistical Mechanics
  and its Applications}\ }\textbf {\bibinfo {volume} {186}},\ \bibinfo {pages}
  {513 } (\bibinfo {year} {1992})}\BibitemShut {NoStop}%
\bibitem [{\citenamefont {Cohen}\ and\ \citenamefont
  {Scully}(1986)}]{Cohen1986}%
  \BibitemOpen
  \bibfield  {author} {\bibinfo {author} {\bibfnamefont {L.}~\bibnamefont
  {Cohen}}\ and\ \bibinfo {author} {\bibfnamefont {M.}~\bibnamefont {Scully}},\
  }\href {\doibase 10.1007/BF01882690} {\bibfield  {journal} {\bibinfo
  {journal} {Foundations of Physics}\ }\textbf {\bibinfo {volume} {16}},\
  \bibinfo {pages} {295} (\bibinfo {year} {1986})}\BibitemShut {NoStop}%
\bibitem [{\citenamefont {Chaturvedi}\ \emph {et~al.}()\citenamefont
  {Chaturvedi}, \citenamefont {Ercolessi}, \citenamefont {Marmo}, \citenamefont
  {Morandi}, \citenamefont {Mukunda},\ and\ \citenamefont {Simon}}]{ref1}%
  \BibitemOpen
  \bibfield  {author} {\bibinfo {author} {\bibfnamefont {S.}~\bibnamefont
  {Chaturvedi}}, \bibinfo {author} {\bibfnamefont {E.}~\bibnamefont
  {Ercolessi}}, \bibinfo {author} {\bibfnamefont {G.}~\bibnamefont {Marmo}},
  \bibinfo {author} {\bibfnamefont {G.}~\bibnamefont {Morandi}}, \bibinfo
  {author} {\bibfnamefont {N.}~\bibnamefont {Mukunda}}, \ and\ \bibinfo
  {author} {\bibfnamefont {R.}~\bibnamefont {Simon}},\ }\href {\doibase
  10.1007/BF02705275} {\bibfield  {journal} {\bibinfo  {journal} {Pramana}\
  }\textbf {\bibinfo {volume} {65}},\ \bibinfo {pages} {981}}\BibitemShut
  {NoStop}%
\bibitem [{\citenamefont {Gross}(2006)}]{gross1.2393152}%
  \BibitemOpen
  \bibfield  {author} {\bibinfo {author} {\bibfnamefont {D.}~\bibnamefont
  {Gross}},\ }\href {\doibase http://dx.doi.org/10.1063/1.2393152} {\bibfield
  {journal} {\bibinfo  {journal} {Journal of Mathematical Physics}\ }\textbf
  {\bibinfo {volume} {47}},\ \bibinfo {eid} {122107} (\bibinfo {year} {2006})}\BibitemShut {NoStop}%
\bibitem [{\citenamefont {Wootters}(1987)}]{WOOTTERS19871}%
  \BibitemOpen
  \bibfield  {author} {\bibinfo {author} {\bibfnamefont {W.~K.}\ \bibnamefont
  {Wootters}},\ }\href {\doibase
  http://dx.doi.org/10.1016/0003-4916(87)90176-X} {\bibfield  {journal}
  {\bibinfo  {journal} {Annals of Physics}\ }\textbf {\bibinfo {volume}
  {176}},\ \bibinfo {pages} {1 } (\bibinfo {year} {1987})}\BibitemShut
  {NoStop}%
\bibitem [{\citenamefont {Wootters}(1986)}]{Wootters1986}%
  \BibitemOpen
  \bibfield  {author} {\bibinfo {author} {\bibfnamefont {W.}~\bibnamefont
  {Wootters}},\ }\href {\doibase 10.1007/BF01882696} {\bibfield  {journal}
  {\bibinfo  {journal} {Foundations of Physics}\ }\textbf {\bibinfo {volume}
  {16}},\ \bibinfo {pages} {391} (\bibinfo {year} {1986})}\BibitemShut
  {NoStop}%
\bibitem [{\citenamefont {Wootters}(2004)}]{Wootters:2004:PQP:1014615.1014629}%
  \BibitemOpen
  \bibfield  {author} {\bibinfo {author} {\bibfnamefont {W.~K.}\ \bibnamefont
  {Wootters}},\ }\href {\doibase 10.1147/rd.481.0099} {\bibfield  {journal}
  {\bibinfo  {journal} {IBM J. Res. Dev.}\ }\textbf {\bibinfo {volume} {48}},\
  \bibinfo {pages} {99} (\bibinfo {year} {2004})}\BibitemShut {NoStop}%
\bibitem [{\citenamefont {Gibbons}\ \emph {et~al.}(2004)\citenamefont
  {Gibbons}, \citenamefont {Hoffman},\ and\ \citenamefont
  {Wootters}}]{PhysRevA.70.062101}%
  \BibitemOpen
  \bibfield  {author} {\bibinfo {author} {\bibfnamefont {K.~S.}\ \bibnamefont
  {Gibbons}}, \bibinfo {author} {\bibfnamefont {M.~J.}\ \bibnamefont
  {Hoffman}}, \ and\ \bibinfo {author} {\bibfnamefont {W.~K.}\ \bibnamefont
  {Wootters}},\ }\href {\doibase 10.1103/PhysRevA.70.062101} {\bibfield
  {journal} {\bibinfo  {journal} {Phys. Rev. A}\ }\textbf {\bibinfo {volume}
  {70}},\ \bibinfo {pages} {062101} (\bibinfo {year} {2004})}\BibitemShut
  {NoStop}%
\bibitem [{\citenamefont {Holmes}\ \emph {et~al.}(2005)\citenamefont {Holmes},
  \citenamefont {Schudy},\ and\ \citenamefont {Aravind}}]{Holmes2005}%
  \BibitemOpen
  \bibfield  {author} {\bibinfo {author} {\bibfnamefont {M.}~\bibnamefont
  {Holmes}}, \bibinfo {author} {\bibfnamefont {W.}~\bibnamefont {Schudy}}, \
  and\ \bibinfo {author} {\bibfnamefont {P.~K.}\ \bibnamefont {Aravind}},\
  }\href@noop {} {\bibfield  {journal} {\bibinfo  {journal} {Thesis:Bachelor of
  science, Worcester Polytechnic Institute}\ } (\bibinfo {year}
  {2005})}\BibitemShut {NoStop}%
\bibitem [{\citenamefont {Gisin}\ and\ \citenamefont
  {Popescu}(1999)}]{PhysRevLett.83.432}%
  \BibitemOpen
  \bibfield  {author} {\bibinfo {author} {\bibfnamefont {N.}~\bibnamefont
  {Gisin}}\ and\ \bibinfo {author} {\bibfnamefont {S.}~\bibnamefont
  {Popescu}},\ }\href {\doibase 10.1103/PhysRevLett.83.432} {\bibfield
  {journal} {\bibinfo  {journal} {Phys. Rev. Lett.}\ }\textbf {\bibinfo
  {volume} {83}},\ \bibinfo {pages} {432} (\bibinfo {year} {1999})}\BibitemShut
  {NoStop}%
\bibitem [{\citenamefont {K.Srinivasan}\ and\ \citenamefont
  {G.Raghavan}()}]{sriniGR}%
  \BibitemOpen
  \bibfield  {author} {\bibinfo {author} {\bibnamefont {K.Srinivasan}}\ and\
  \bibinfo {author} {\bibnamefont {G.Raghavan}},\ }\href {\doibase
  arXiv:1604.00188} {\ arXiv:1604.00188}\BibitemShut {NoStop}%
\end{thebibliography}

%

\end{document}